\documentclass[aps,prl,twocolumn,groupedaddress,showpacs,showkeys,floatfix]{revtex4-1}
\usepackage[dvipdfm]{graphicx, color}
\usepackage{colordvi}
\usepackage{color}
\usepackage{float}
\usepackage{amsmath}
\usepackage{amssymb}
\usepackage{comment}
\usepackage{booktabs}

\usepackage{lineno}

\begin{document}
\title{Slow Stochastic Switching by Collective Chaos of Fast Elements}

\author{Hidetoshi Aoki}
\author{Kunihiko Kaneko}
\affiliation{
Research Center for Complex Systems Biology,
Graduate School of Arts and Sciences,
The University of Tokyo 3-8-1 Komaba, Meguro-ku Tokyo 153-8902, Japan}
\date{\today}
\begin{abstract}
Coupled dynamical systems with one slow element and many fast
elements are analyzed. By averaging over the dynamics of the fast
variables, the adiabatic kinetic branch is introduced for the
dynamics of the slow variable in the adiabatic limit. The dynamics
without the limit are found to be represented by stochastic
switching over these branches mediated by the collective chaos of
the fast elements, while the switching frequency shows a complicated
dependence on the ratio of the two timescales with some resonance
structure. The ubiquity of the phenomena in the slow--fast dynamics
is also discussed.
\end{abstract}

\maketitle

Dynamics with distributed timescales are ubiquitous in nature, not
only in physicochemical and geophysical systems but also in
biological, neural, and social systems. In biological rhythms, for
example, dynamics with timescales as long as a day coexist and
interfere with the dynamics of much faster biochemical reactions
occurring on subsecond timescales \cite{Winfree}. A similar
hierarchy exists even within protein dynamics \cite{protein}.
Electroencephalography (EEG) of the brain is known to involve a
broad range of frequencies, and the functional significance of
multiple timescales has been extensively discussed
\cite{brain1,brain2,brain3}: Neural dynamics in higher cortical
areas alter our attention on a slower timescale and switch the
neural activities of faster timescales in lower cortical areas.
Faster sensory dynamics are stored successively in short-term to
long-term memory. Unveiling the salient intriguing behavior that is
a result of the interplay of dynamics with different timescales is
thus of general importance.

To treat dynamics with fast and slow timescales, several theoretical
tools have been developed since the proposition of Born--Oppenheimer approximation
in quantum physics. Consider dynamical systems of the
form
\begin{equation}
 dy_i/dt=F_i(\{x_j,y_j\}) ;\;\; \epsilon dx_i/dt=G_i(\{x_j,y_j\}),
\end{equation}
where $\epsilon$ is small so that $\{ x_i \}$ are faster variables
than $\{ y_j \}$. According to adiabatic elimination or Haken's slaving
principle \cite{Haken,GH,KK81,singular}, fast variables are
eliminated by solving $dx_i/dt=0$ for a given $\{ y_j \}$, and by
using this solution of $\{ x_i \}$ as a function of $\{ y_j \}$,
closed equations for the slow variables are obtained. This is a
powerful technique when the fast variables are relaxing to fixed
points for the given slow variables, whereas to include a case for
which the fast variables have oscillatory dynamics, the averaging
method is useful \cite{GH,Average}. That is, the long-term average
of the fast variables $<x_i>$ is taken for a given $\{ y_j\}$, and
by inserting the average into the equation for $\{ y_j \}$, a set of
closed equations for the slow variables is obtained. When the number
of variables involved is small, additional techniques developed with
the use of a slow manifold can be beneficial \cite{Guckenheimer}.
Dynamical systems with mutual interference between the fast and slow
variables have also been investigated
\cite{Fujimoto1,Kanz,Fujimoto2,Vulpiani,synchro,Tachikawa,Spain}.

In this Letter, we study a case that involves a large number of fast
variables which show chaotic dynamics. We introduce the adiabatic
kinetic plot (AKP) to account for the kinetics of the slow variables
under the adiabatic limit $\epsilon \rightarrow 0$ by using the
averaging method. We show that this plot is useful for analyzing the
dynamics even for a finite, small $\epsilon$ for which stochastic
transitive dynamics over different modes are observed and are
explained as switches over the adiabatic kinetic branches (AKB)
obtained from the AKP. This stochasticity in the switches is shown
to originate from the collective chaos of an ensemble of fast
variables.

\begin{figure}[htbp]
\includegraphics[width=4cm,height=4cm]{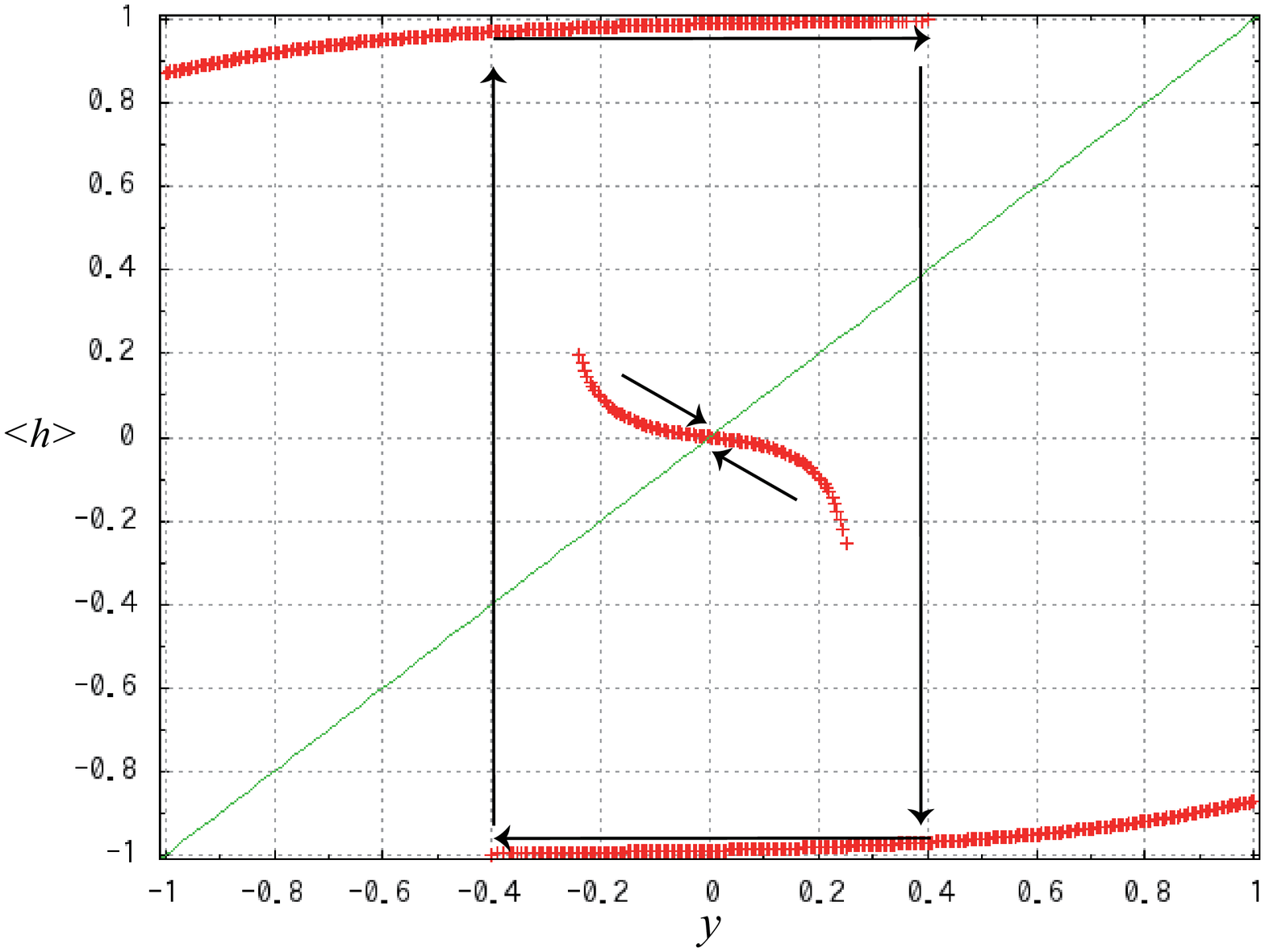}(a)
\includegraphics[width=7cm,height=3cm]{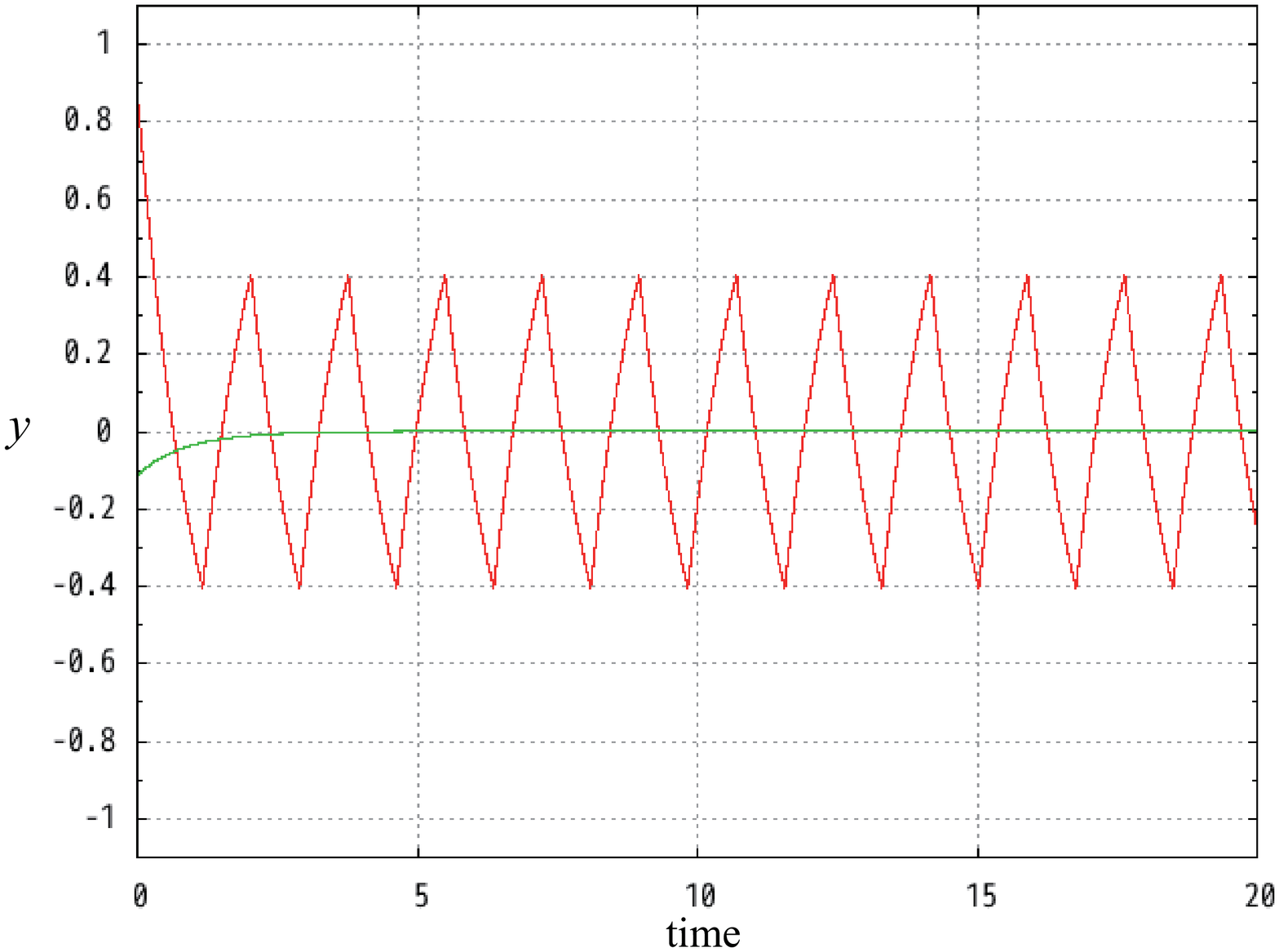}(b)
\caption{
(Color online). (a) An example of the AKP for our model (eq.(1)(2))
with $N=10$ and $\beta=10$. The AKP is computed by fixing the $y$ value at the
abscissa and numerically integrating the equations of the fast
variables for $10^3$ time units from 10 randomly selected initial
conditions. The ordinate $<h>$ is then computed from the temporal
average of the last $8\times 10^2$ time units of each initial
condition. The $y$ value is incremented by $0.005$ to obtain the
plot. (b) The time series of $y$ for $\epsilon=0.0001$. There are
two coexisting attractors: a fixed point (green) and a limit cycle
(red).
}
\label{fig:1}
\end{figure}

As a specific example, we consider the case for a single slow
variable $y$, where $F\equiv h(\{x_j\},y) -y$ and $G(\{x_j\},y)$ are chosen from the threshold
dynamics as
\begin{equation}
\frac{dy}{dt}= h(\{x_j\},y) - y \equiv tanh (\frac{\beta}{\sqrt{N}}
\sum_{j=1}^{N-1}(J_{0j}x_j + J_{00}y)) - y,
\end{equation}
\begin{equation}
\epsilon \frac{dx_i}{dt}=
G(\{x_j\},y) \equiv tanh ( \frac{\beta}{\sqrt{N}} \sum_{j=1}^{N-1}
(J_{ij}x_j+J_{i0}y) )- x_i,
\end{equation}
where $\beta$ is taken to be 10. Here, $J_{ij}$ is chosen as a
homogeneous random number in the interval [$-$1,1], and once it is
chosen, it is fixed during the dynamics for each sample. We have
adopted this form because this type of threshold dynamics
\cite{thr,thr2} is used as a simplification of neural network
\cite{thr-NN} or gene regulation network dynamics \cite{thr-GRN}, in
which each element (neuron or gene expression) tends to take either
an ``on'' ($x=1$) or ``off'' ($x=-1$) state activated ($J_{ij}>0$)
or inhibited ($J_{ij}<0$) by other elements. Note, however, that the
method and findings discussed here are not restricted to the
specific choices of the functions $F$ and $G$; they are valid for
any choice.

The dynamics of the slow variable $y$ is represented using the
averaging method as $dy/dt=< h(\{x_j\}(y))>-y,$ where
$<h(\{x_j\}(y))>$ is the temporal average of $h$ for a given $y$,
i.e., the average input that $y$ receives from $\{x_j \}$, in the
adiabatic limit. To compute the average $<\cdot>$, we first fix the
$y$ value, obtain the attractors for $x_j$, and then compute the
temporal average for each attractor. By changing the value of $y$,
$<h(\{x_j\}(y)>$ is obtained, and this forms a continuously changing
branch. At this point, it is useful to introduce the plot $(y,<h(\{x_j\}(y))>)$ (Fig.~\ref{fig:1}) If there are multiple
attractors that depend on the initial condition of $x_j$, then there
are several branches in the AKP. Starting from a given $y$ and
initial condition $x_j$, the dynamical system falls on a specific
branch. According to the equation for $y$, if $<h>$ is larger
(smaller) than $y$, then $dy/dt>0$ ($dy/dt<0$). Thus, we can trace
the dynamics of $y$ along each branch. When a branch crosses the
line $y=<h>$, then $y$ falls on a fixed point. If the slope of the
branch at the fixed point is less than unity, then the system is attracted to
the point so that the slow variable falls on a fixed point
attractor (at least) in the limit of $\epsilon \rightarrow 0$ (see
the middle branch in Fig.~\ref{fig:1}). We have confirmed that this
is true up to a certain value of $\epsilon$.

The periodic motion of $y$ can also be explained by the AKP. For example, see the top and
bottom branches in Fig.~\ref{fig:1}. As $y$ increases along the top
branch, it eventually reaches the endpoint of the branch and then
switches to the bottom branch, which corresponds to an alternative
attractor of $x$. The process then repeats itself as $y$ decreases
along the bottom branch to the endpoint before switching to the top
branch. Indeed, this periodic oscillation exists as an attractor, as shown in
Fig.~\ref{fig:1}(b). In this example, the $\{ x_j \}$ attractor is a
fixed point at each branch, but in many other examples, the
attractor may be a limit cycle or chaos. However, the present
analysis of the $y$ dynamics is still valid in such cases.
In fact, the periodic oscillation of $y$ as analyzed from the AKP
exists up to a certain value of $\epsilon$ (e.g., $\sim$0.01), where
a small amplitude, fast oscillation of order $\epsilon$ is added to
the slow $y$ oscillation, if $\{ x_j\}$ exhibits oscillation.

\begin{figure}[t]
\begin{center}
\includegraphics[width=8cm]{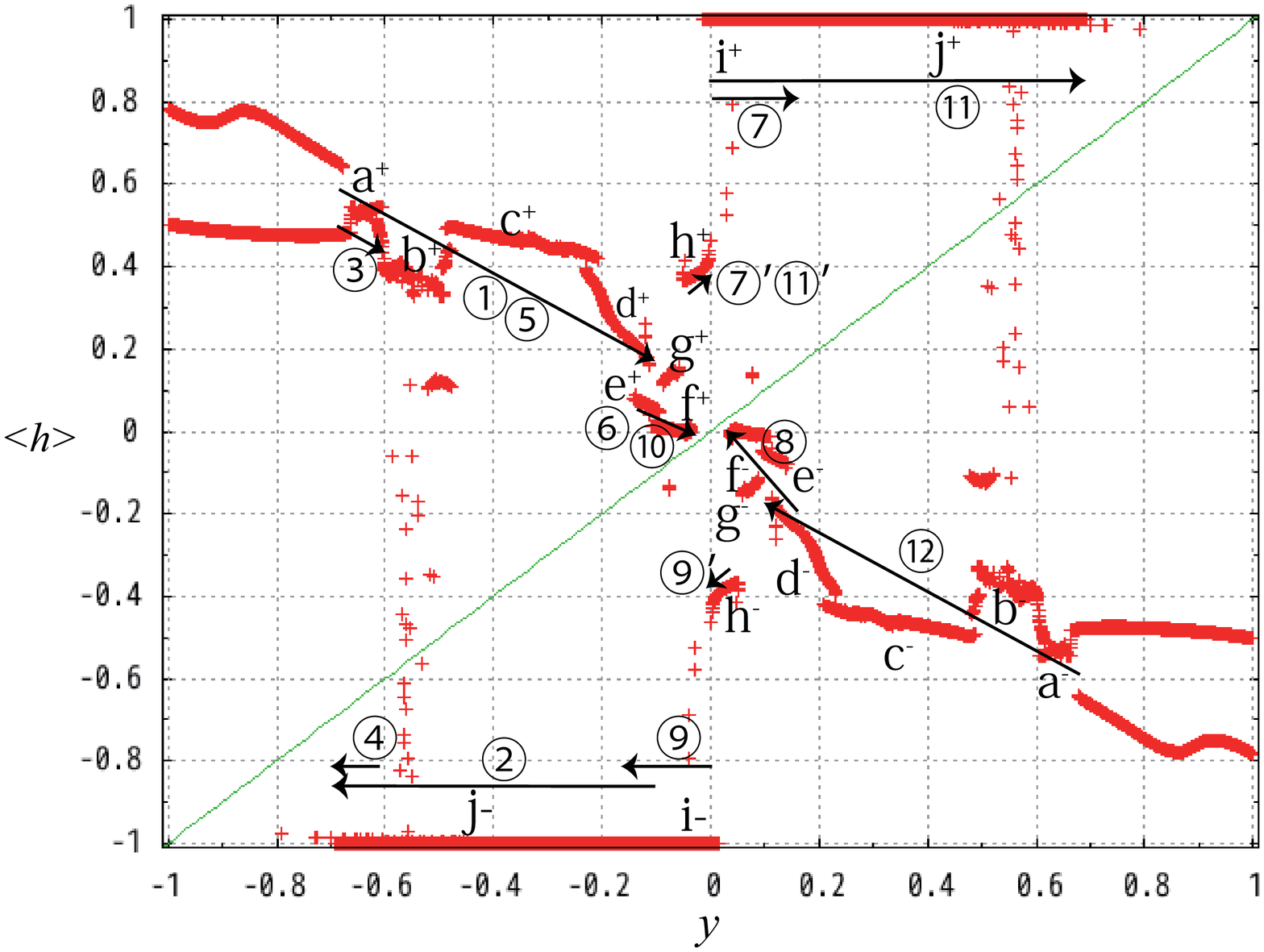}(a)
\includegraphics[width=8cm,height=4cm]{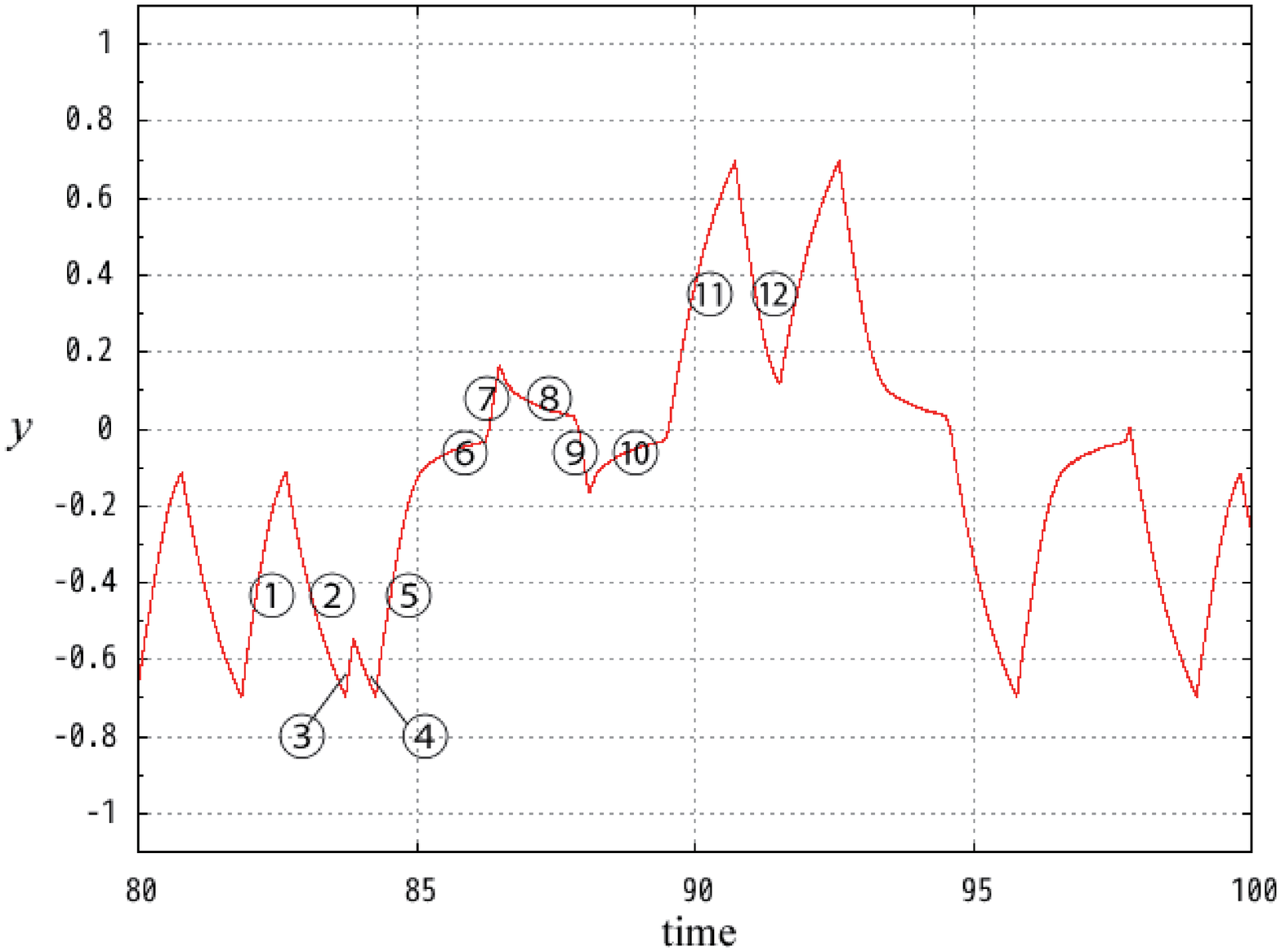}(b)
\begin{tabular}{c} 
\begin{minipage}{0.33\hsize}
    \includegraphics[width=2.8cm,height=2.6cm]{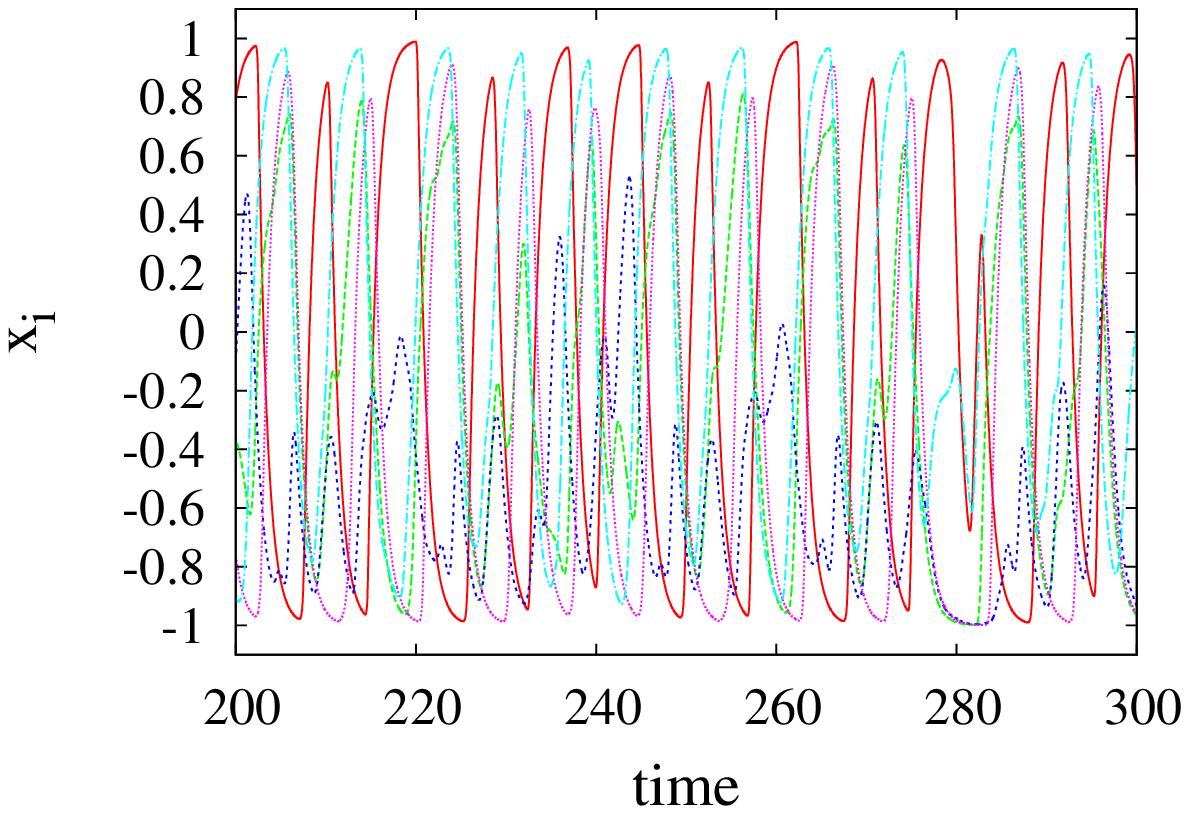}(c1)
\end{minipage}
\begin{minipage}{0.33\hsize}
    \includegraphics[width=2.8cm,height=2.6cm]{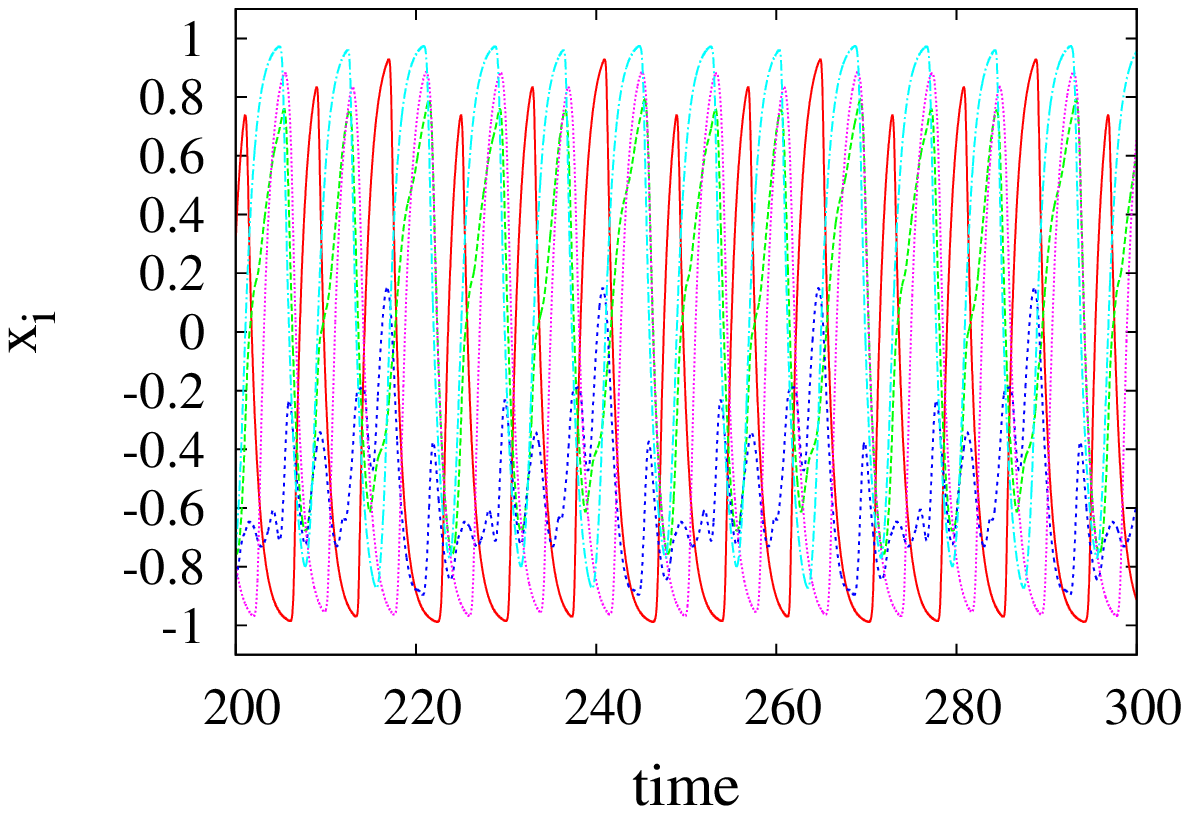}(c2)
\end{minipage}
\begin{minipage}{0.33\hsize}
    \includegraphics[width=2.8cm,height=2.6cm]{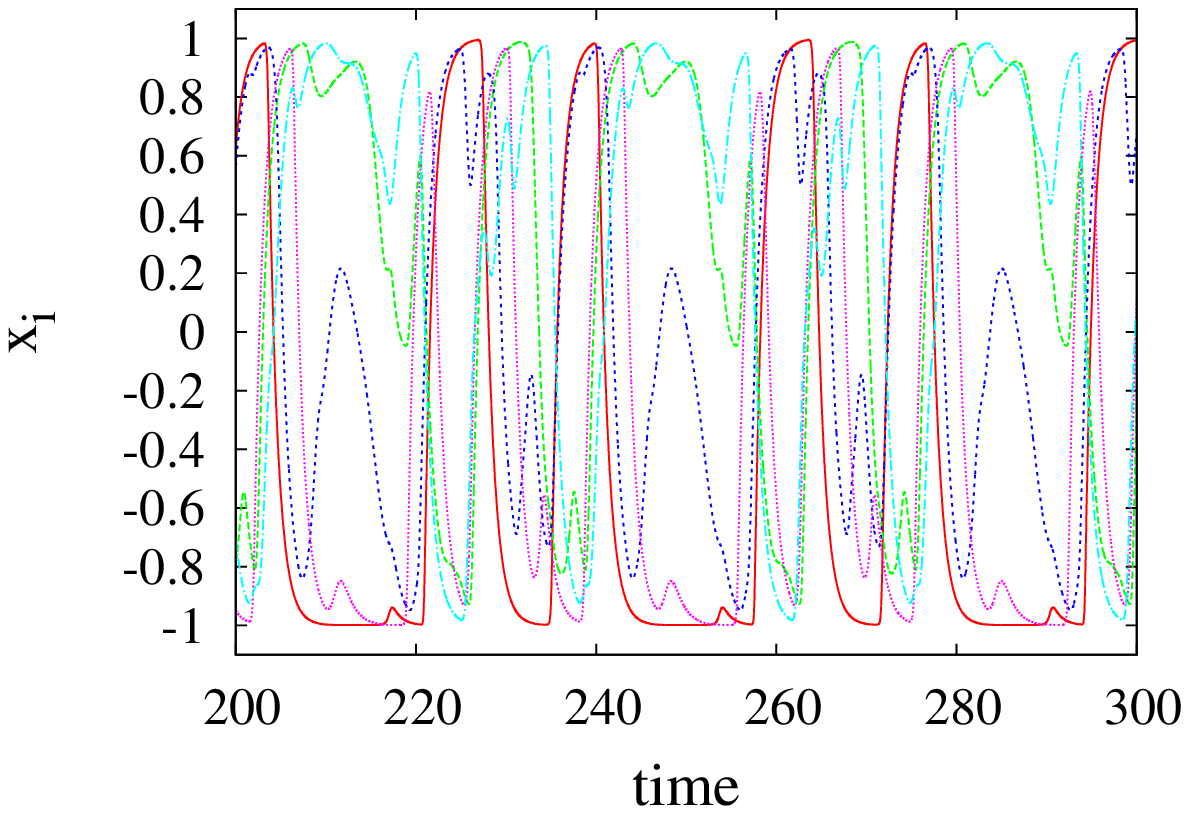}(c3)
\end{minipage}
\end{tabular}
\caption{(Color online) (a) The AKP with $N=30$ for a given $J$ that
produces complex oscillation. The AKP is computed in the same
manner as Fig.~\ref{fig:1} but from $3\times 10^5$ to $3.5 \times
10^5$ time units over 20 randomly chosen initial conditions. The
branches are labeled with lower case letters. Due to the symmetry in
the model, branches for $y>0$ and $y<0$ are indicated by $+$ and
$-$, respectively. (The branches $i$ and $j$ happen to take almost
the same $<h>$ value but belong to different $\{ x_j \}$
attractors). The numbers correspond to the time course plotted in part (b).
(b) Time series of the slow variable $y$ for $\epsilon=0.0001$.
The numbers (1, 2, $\cdots$, 12) correspond to the branches visited there,
as displayed in part (a). 
(c) The time series of the fast variables $x_i$
($i=1,2,\cdots,9$) for the branches b, c, and h. 
}
\label{fig:2}
\end{center}
\end{figure}

\begin{figure}[t]
\begin{center}
\begin{tabular}{c} 
\begin{minipage}{0.33\hsize}
\includegraphics[width=2.8cm,height=2.6cm]{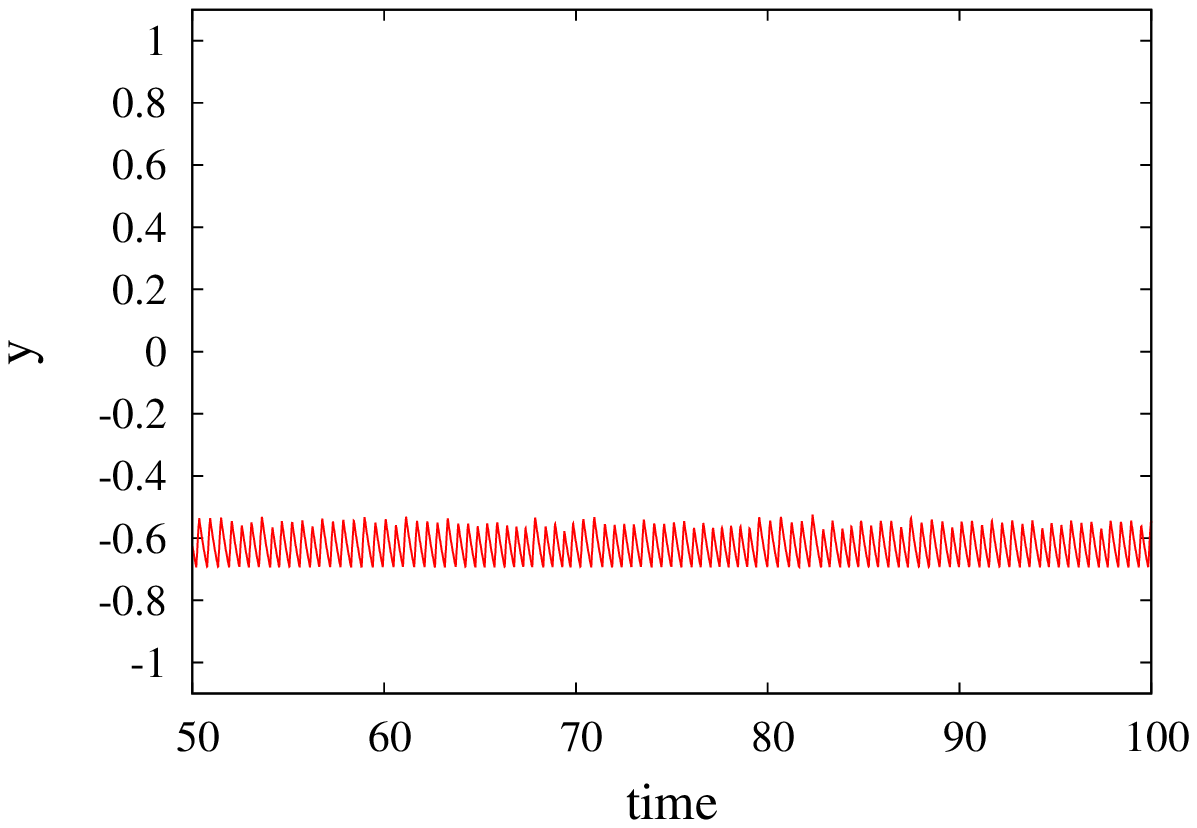}(a)
\end{minipage}
\begin{minipage}{0.33\hsize}
\includegraphics[width=2.8cm,height=2.6cm]{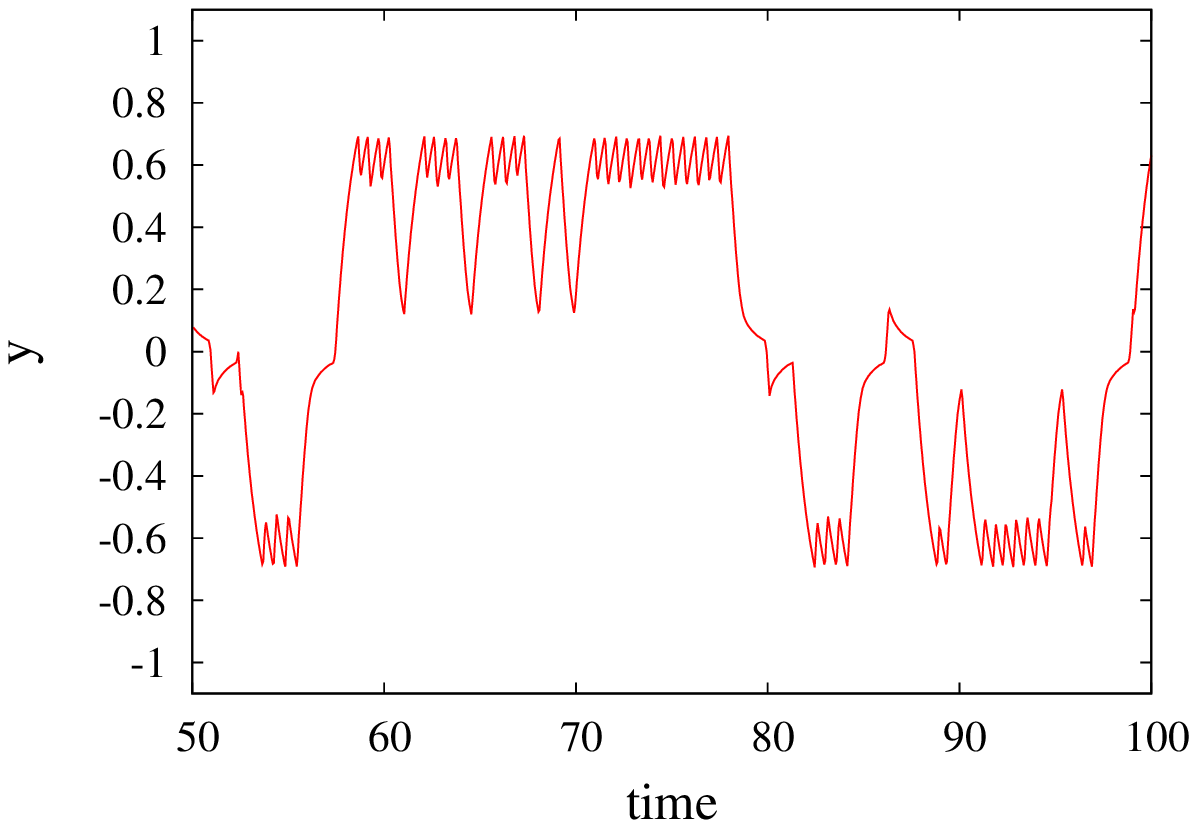}(b)
\end{minipage}
\begin{minipage}{0.33\hsize}
\includegraphics[width=2.8cm,height=2.6cm]{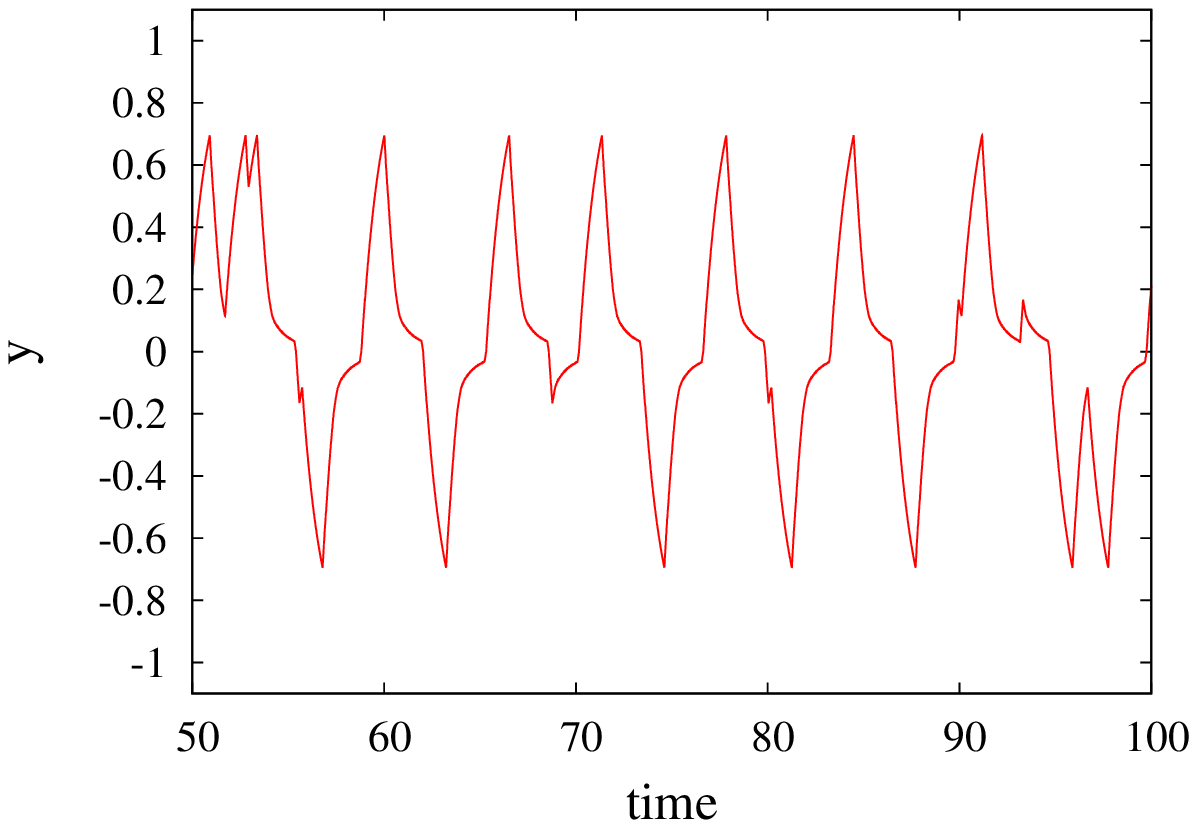}(c)
\end{minipage}
\end{tabular}
\caption{(Color online): Time 
series of the slow variables $y$ corresponding to Fig,2, except for $\epsilon=2\times 10^{-6}$ (a),
$\epsilon=1\times 10^{-5}$ (b), and $\epsilon=1\times 10^{-4}$
(c). The time series of (c) corresponds to Fig.2(b) but is plotted
for a longer time span.}  \label{fig:25}
\end{center}
\end{figure}

In general, AKP has much more branches that make the oscillatory
dynamics complex. A rather more complicated example is shown in
Fig.~\ref{fig:2}. In this case, in the limit of $\epsilon
\rightarrow 0$, $y$ switches between two branches. In the example in
Fig.~\ref{fig:25} (a) for $\epsilon=2\times 10^{-6}$, $y$
periodically switches between the branch $a^{+}$ and a section of
branch $j^{-}$ (``4''). For a larger $\epsilon$ value, however,
complex oscillations of $y$ are seen, as shown in Fig.~\ref{fig:2}. 
This is described as the switching over all 2 $\times$ 10
branches, a$^{\pm}$, b$^{\pm}$, c$^{\pm}$, ..., j$^{\pm}$ (the first
12 are labeled explicitly in the figure), where $\pm$ denotes the
symmetric branches of $y>0$ and $y<0$. Here, we should mention that
this switching is not always deterministic. For example, $a^{+}
\rightarrow b^{+}$ (``1'') or $a^{+} \rightarrow i^{-}$ (``3''
$\rightarrow$ ``4'') are both possible, as are $d^{+} \rightarrow
j^{-}$ (``1'' $\rightarrow$ ``2'') and $d^{+} \rightarrow e^{+}$
(``5'' $\rightarrow $ ``6''). As $\epsilon$ is decreased, a larger
number of branches is visited by the stochastic switches
(see Fig.~\ref{fig:25}(a) to \ref{fig:25}(b) and to \ref{fig:25} c)] 
until only a cycle between two branches remains in the
limit of $\epsilon \rightarrow 0$, as in Fig.~\ref{fig:25}(a).

With the complex switches, the dynamics of $x_i$ switch among (at
least) 2 $\times$ 10 types of attractors including fixed points,
limit cycles, and chaos. This type of switching is reminiscent of
chaotic itinerancy \cite{Ikeda,Tsuda,GCM,CI} where the orbit
itinerates over ``attractor ruins''. Here, in contrast, the
stochastic switches progress among attractors for a given value of
the slow variable $y$, while the chaotic dynamics of the fast
variables provides a source for the stochastic switching. Indeed, at
the boundary of the branches $a,d,f,\cdots$, the fast variables
$\{x_i \}$ show chaotic oscillation.

\begin{figure}[!ht]
\begin{center}
\includegraphics[width=5cm,height=5cm]{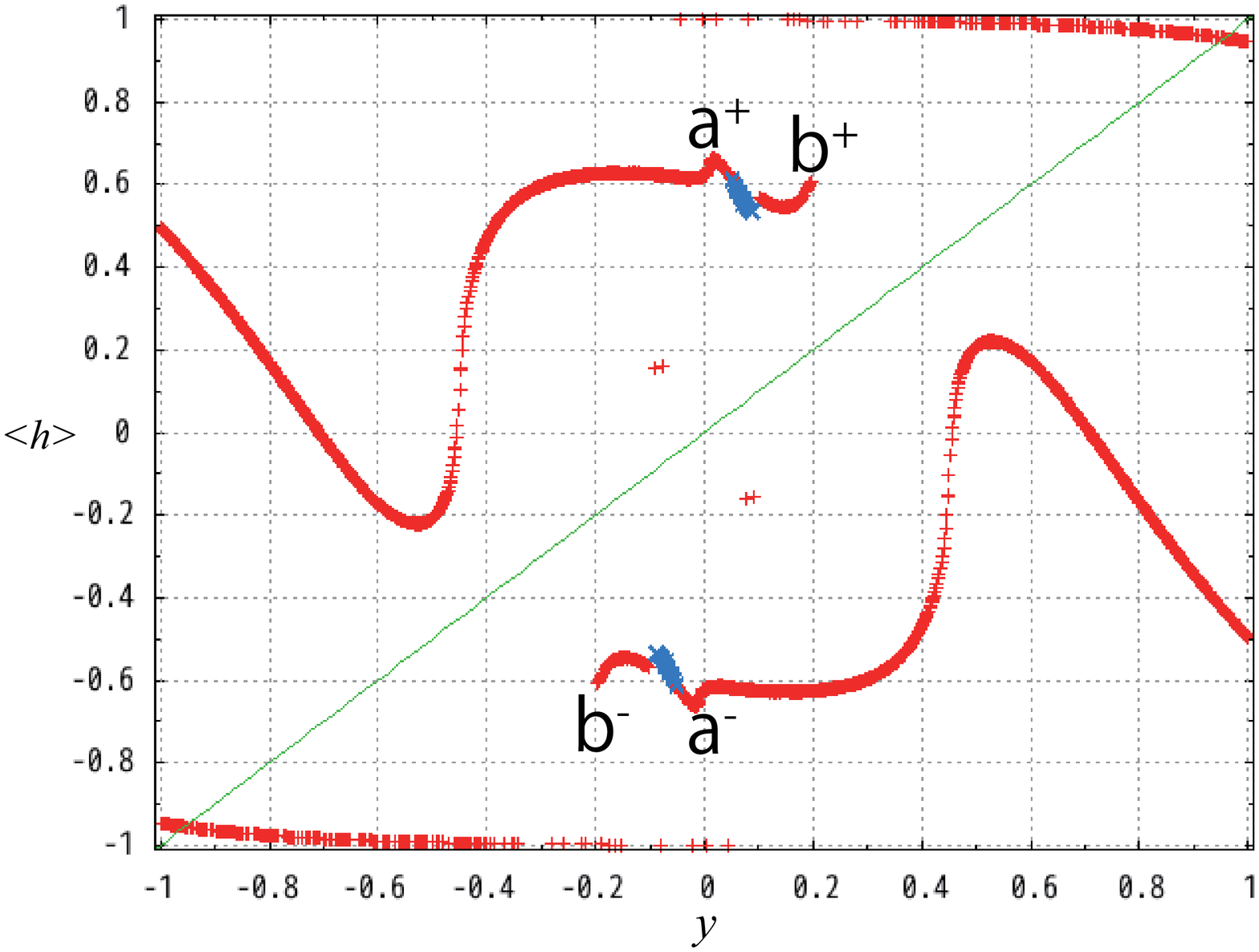}(a)
\end{center}
\begin{center}
\includegraphics[width=8cm,height=4cm]{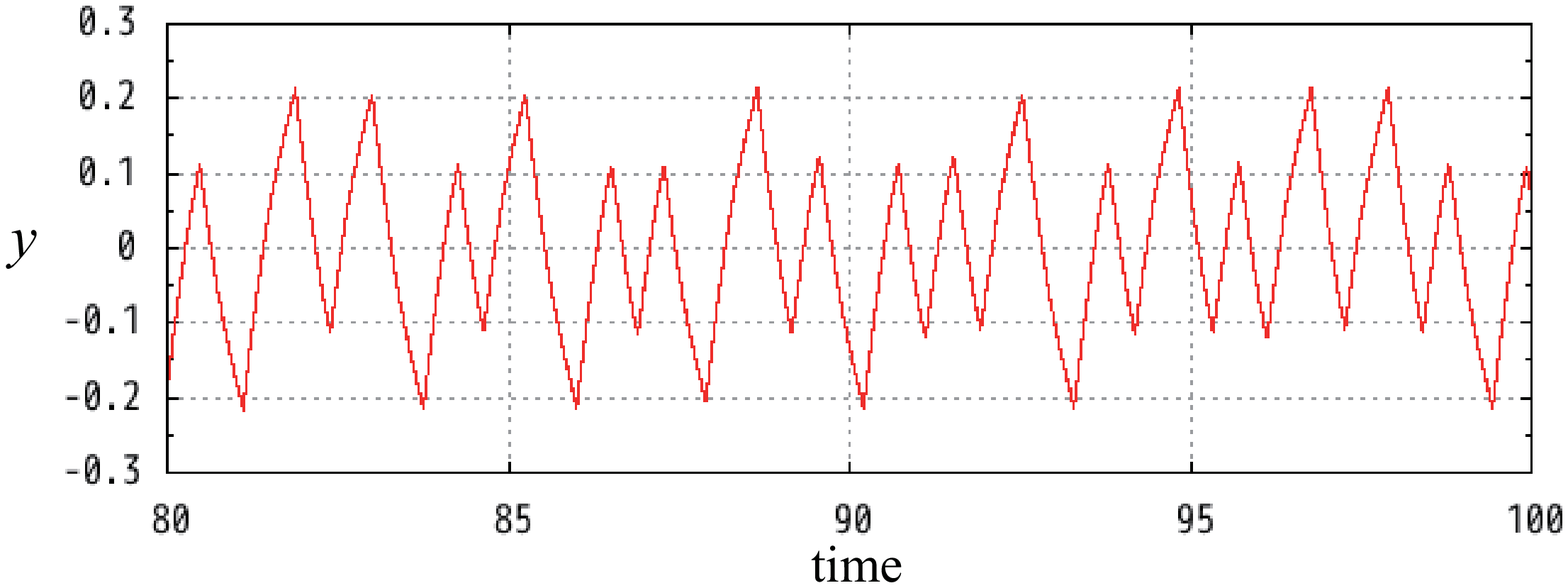}(b)
\end{center}
\caption{(Color online) (a) The AKP computed for a different matrix
$J$ with $N=20$ and the time average between $10^2$ to $2\times
10^3$ time units from 15 initial conditions. 
The maximum Lyapunov exponent of the
dynamics of $\{ x_i \}$ is also computed over the interval of $3\times 10^3$
to $4\times 10^3$ time units for each branch of a given $y$. The segment of the branch with
positive exponent (whose value is about $\sim .02$) is colored as blue. (See Supplementary Fig.1 for
the Lyapunov exponents at each branch).  
(b) The time series of $y$ for $\epsilon=0.001$. 
}
\label{fig:3}
\end{figure}

For a detailed analysis of the stochastic switching due to the
chaotic dynamics, we consider the simpler example given in
Fig.~\ref{fig:3} with a different matrix $J_{ij}$. In this case, as
$\epsilon \rightarrow 0$, $y$ shows periodic oscillation between the
two branches $a^{+}$ and $a^{-}$, whereas for $\epsilon> \epsilon_c
\approx 0.5 \times 10^{-5}$, the branches $b^{+}$ and $b^{-}$ are also
available, and stochastic switching $a^{+} \rightarrow a^{-}, b^{+}$
and its symmetric counterpart appear. The choice between $a^{+}
\rightarrow a^{-}$ and $a^{+} \rightarrow b^{+}$ is stochastic, and
indeed, we have computed the Shannon entropy of the $n$-tuple symbol sequence of the
branches $a,b$ visited by the slow $y$ variable and confirmed that it increases
linearly with $n$ ($ \sim 0.91n$)\cite{transient}.

To examine if the origin of the stochasticity lies in the chaos of
the fast variables, we measured the maximal Lyapunov exponent for
the $(N-1)$-dimensional fast dynamics of $\{x_i \}$ for a given $y$
at each branch. As shown in Fig.~\ref{fig:3} (c), the exponent is
positive around the endpoints of the branches where stochastic
switching occurs. Several other examples
also show stochastic transitioning beyond a critical value of
$\epsilon$, a Poisson switching-time distribution, and a positive
Lyapunov exponent at the branch endpoint 
(for example see Supplementary Figs. 2). 
The stochastic switching
from the branches a and f in Fig.~\ref{fig:2} is also explained by the
chaotic dynamics of the fast $\{x_i \}$ variables at the branches.

\begin{figure}[h]
\includegraphics[width=6cm,height=4cm]{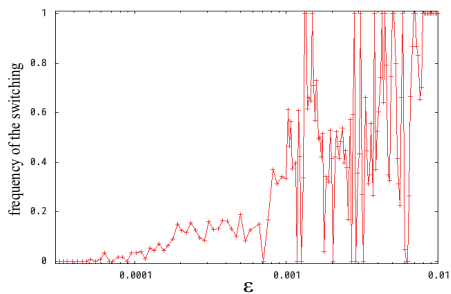}
\caption{(Color online) Frequency of the switching in the branches
corresponding to the dynamics in Fig.~\ref{fig:3} as a function of
$\epsilon$. For each $\epsilon$ value, the number of a+ to b+
switching events divided by those from a+ to a- are computed as the
fraction of the number of times that $y$ goes beyond $0.2$ divided
by the number of times that it goes below $0$ within 500 time units.
} \label{fig:4}
\end{figure}

When $\{ x_i \}$ shows chaotic dynamics, one might expect that the variable $h$ can be regarded as just the
noise represented by the sum of random $\{ x_i \}$ variables. If
this were the case, then the amplitude of this noise would decrease
as the number of fast elements $N$ is increased. The variable $h$ would
then approach a constant in the $N\rightarrow\infty$ limit, and the
frequency of the stochastic switchings would decrease accordingly.
However, this is does not occur. We simulated the present model by
increasing the number of fast elements by $2^k$ ($k=1,2,\cdots ,6$)
by cloning the matrix $J_{ij}$ and confirmed that the frequency does
not decrease with an increase in $N$. This suggests that there is
still some correlation among the fast variables $x$, so that $h$
shows collective chaotic motion, as has been studied extensively
\cite{KK-PRL,CM,Shibata,KK-book}. In fact, the oscillation of the
collective variable $h$ has a larger amplitude than the typical
mean-field oscillation in the collective chaos in coupled chaotic
systems studied thus far \cite{KK-PRL,Shibata}. Indeed,
some of the fast elements $x_i$ undergo a large-amplitude change
between the on and off (1 and $-$1) states and there remains
correlation among the $x_i$ variables.

With an increase in $\epsilon$ beyond $\epsilon_c$, the frequency of
the stochastic switching increases, but with a further increase, the
frequency shows a complicated dependence on $\epsilon$, as it is increased beyond $\epsilon _c$.
There are certain parameters for which the switching
loses its stochasticity and is replaced by either perfect switching
between two original branches in the $\epsilon\rightarrow 0$ limit
or perfect switching to the new branch (i.e., $a^+ \rightarrow b^+$
only). When the switching ratio is zero or unity, the long-term
oscillation of the slow variable $y$ is periodic. Thus, each such
``deterministic'' region is regarded as a ``window'' in the
parameter region showing chaos. Here, it is interesting to note that
periodic motion is generated between variables with timescales
differing by more than one order of magnitude. In fact, the
collective variable $h$ can have a slower component than the
original timescale $\epsilon$ for $x_i$. Complicated resonance
structures of the switching ratio are often observed in the present
system when stochastic switchings exist.

To summarize, we have introduced an AKP to study the kinetics of
slow variables in the adiabatic limit $\epsilon \rightarrow 0$, 
Up to a certain critical value $\epsilon_c$, the slow dynamics fall either on a
fixed point or exhibit periodic switching between branches. As
$\epsilon$ is increased (i.e., the timescale difference is
decreased), stochastic switching among several branches appears
mediated by the collective chaotic motion of the fast variables, and
the variety of switchings increases with a further increase in
$\epsilon$.

Although we have employed a simplified threshold dynamics model that
borrows concepts from neural or gene regulation networks, the AKP
method can be applied generally to fast--slow systems, and
stochastic switching over the AKB will appear when the fast
variables show chaotic motion. Extension to a case with multiple
slow variables is also possible, in principle, by extending each
branch to a surface or higher-dimensional manifold. Although
visualization in this case will be difficult in comparison with the present AKP,
the stochastic transitions over adiabatic manifolds can be
analyzed using the methods developed here.

As a result of the switching, long-term itinerancy over different
modes of oscillation of the fast variables appears, which is
reminiscent of chaotic itinerancy. Experimentally, such itinerancy
is often observed in EEGs of the brain, biorhythms, climate
dynamics, and so forth, where modes with different timescales
coexist \cite{CI}. The present approach may shed light on such
itinerant behavior, while hierarchical construction of AKPs may be beneficial to
deal with a system with a variety of distinct timescales
\cite{Fujimoto2}.

Collective chaotic motion of fast variables is a source of
stochastic switching and is modulated by the motion of slow
variables; such mutual inference between fast and slow modes leads
to resonance between the slow and collective modes, which is similar
to the interference in the neural activity dynamics between higher
and lower cortical areas during changes in our attention.

We would like to thank Shuji Ishihara, Nen Saito, and Shin'ichi Sasa
for useful discussions. This work was partially supported by a
Grant-in-Aid for Scientific Research (No. 21120004) on Innovative
Areas ``Neural creativity for Communication'' (No. 4103) and the
Platform for Dynamic Approaches to Living System from MEXT, Japan.

\bibliographystyle{unsrt}

\begin{thebibliography}{99}
\bibitem{transient}
For some parameter values, the stochastic switching may not continue forever, and the
slow variables are ultimately settled down to periodic motion between a pair of branches.
Possibility in very long transient motion is also common to most examples in chaotic itinerancy.
\end{thebibliography}

\end{document}